\documentclass[prl,reprint]{revtex4-2}
\usepackage{graphicx}
\usepackage{dcolumn}
\usepackage{bm}
\usepackage[svgnames]{xcolor}
\usepackage{mathtools}
\usepackage{dsfont}
\usepackage{url}
\usepackage{hyperref}
\usepackage[capitalise]{cleveref}
\usepackage{braket}
\usepackage[version=4]{mhchem}
\usepackage{ragged2e}
\usepackage{subcaption}
\usepackage{comment}
\usepackage{amssymb}
\usepackage{mathrsfs}
\usepackage{xcolor}
\usepackage{physics}
\usepackage{subfiles}

\usepackage{ulem}
\usepackage{soul}

\begin{document}

\preprint{APS/123-QED}

\title{Finite and disordered Kitaev chains: a large deviation study}

\author{Cl\'ement Fortin$^1$}
\author{Kai Wang$^{1}$}
\author{T. Pereg-Barnea$^{1}$}

\affiliation{$^1$Department of Physics,  Regroupement qu\'{e}b\'{e}cois sur les mat\'{e}riaux de pointe and McGill Quantum Center, McGill University, Montr\'eal, Qu\'ebec H3A 2T8, Canada }

\date{\today}

\begin{abstract}
    Topological edge states are celebrated for their robustness against disorder, yet the interplay between disorder and system size remains poorly understood. We use large deviations theory as a framework to study finite-size effects beyond the central limit theorem. We analyze Lyapunov exponent fluctuations in the static and periodically driven disordered Kitaev chain and find an asymmetry in the large deviations statistics that makes stronger edge localizations of Majorana zero modes exponentially more likely than weaker ones. We demonstrate that this fluctuation asymmetry is not tied to the topological phase. This asymmetry endows topological edge states with an additional protection against disorder and persists across a broad class of disorder distribution. We show how to use our framework to find the minimum system size required to satisfy topological quantum computing constraints.
\end{abstract}

\maketitle

\textit{Introduction}.---Finite systems often exhibit disorder and boundary effects that make them distinct from their infinite-size or periodic counterparts. Despite this discrepancy, it is standard practice to assume that finite systems are large enough to approximate the thermodynamic limit with Gaussian fluctuations. Such an approach is ubiquitous in disordered topological systems because the topological invariant is only meaningfully quantized in the thermodynamic limit \cite{Hasan2010,Qi2011}. This is particularly true in topological materials, where samples engineered for topological quantum computations are riddled with impurities \cite{Sarma2023v2}. However, while modern platforms such as semiconductor-superconductor nanowires may be large enough to exhibit measurable topological signatures \cite{Deng2012,Mourik2012,Churchill2013,Finck2013,Nichele2017,Song2022,Aghaee2023,Aghaee2025}, the features observed can be topologically trivial due to disorder \cite{Pan2020,Pan2020v2,Yu2021,Pan2021,Sarma2023,Pan2024}. 
Moreover, the complexity of disordered systems can make it challenging to benchmark them numerically and assess how their properties evolve with system size. Accurately characterizing the distribution of fluctuations as a function of system size, thereby quantifying the approach to the thermodynamic limit, is thus of central relevance.


In this work, we investigate finite-size effects of topological edge modes and their approach to the thermodynamic limit. We focus on the Majorana edge modes present in the static and driven disordered fermionic Kitaev chains \cite{Kitaev2001,Ling2024}. The transfer matrix formalism has been widely employed to characterize the impact of disorder on Majorana modes and their experimental signatures \cite{Brouwer2011,Sau2012,Ryu2012,Ludwig2013,Fregoso2013,deGottardi2013,Mandal2016,Hedge2016}. However, these studies have focused mainly on disorder-averaged quantities and variances, and the statistics of disorder realizations have not been systematically explored. Using the rate function of large deviations \cite{Bahadur1960,Dembo1998,Hollander2000,Touchette2009}, we analyze the full distribution of Lyapunov exponents $L_N$ of Majorana edge modes across disorder realizations in chains of finite size, $N$. The left edge mode wavefunction depends exponentially on $n$, $\psi_n\sim e^{nL_N}$, where $L_N<0$ corresponds to exponential localization on the left boundary and $L_N>0$ to an Anderson-localized mode. The Lyapunov exponent here is minus the inverse of the localization length. Within a Gaussian, central limit approximation, fluctuations of the Lyapunov exponent $L_N$ about its disorder-averaged value $L_\infty =\expval{L_N}$ are symmetric, where we write $\expval{\cdot}$ to denote disorder averaging. Indeed, small deviations of equal magnitude above and below $L_\infty$ occur with equal probability when the system size becomes large enough. We show that this symmetry can be absent at the level of large deviations.
Fluctuations that favor stronger boundary localization may be exponentially more likely than more extended profiles, providing an additional robustness to disorder that is \textit{distinct} from conventional topological protection.  

\begin{figure*}[ht]
    \centering
    \begin{subfigure}{0.31\textwidth}
        \centering
        \includegraphics[width=\textwidth]{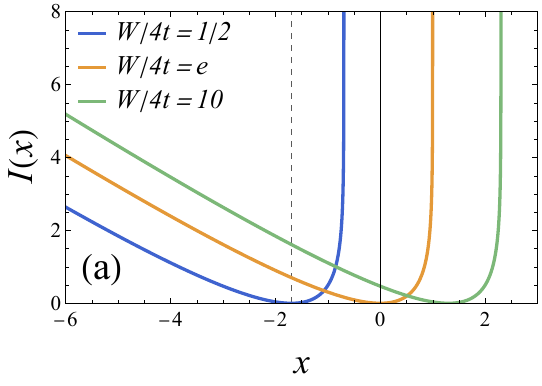}
    \end{subfigure}
    \begin{subfigure}{0.335\textwidth}
        \centering
        \includegraphics[width=\textwidth]{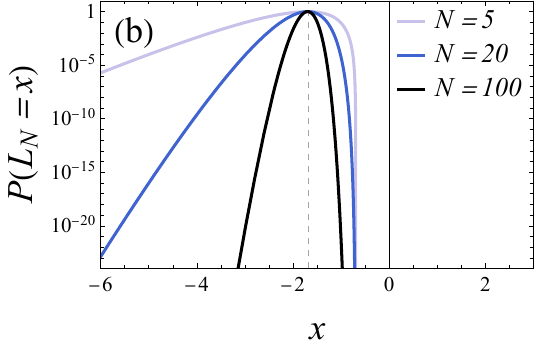}
    \end{subfigure}
     \begin{subfigure}{0.335\textwidth}
        \centering
        \includegraphics[width=\textwidth]{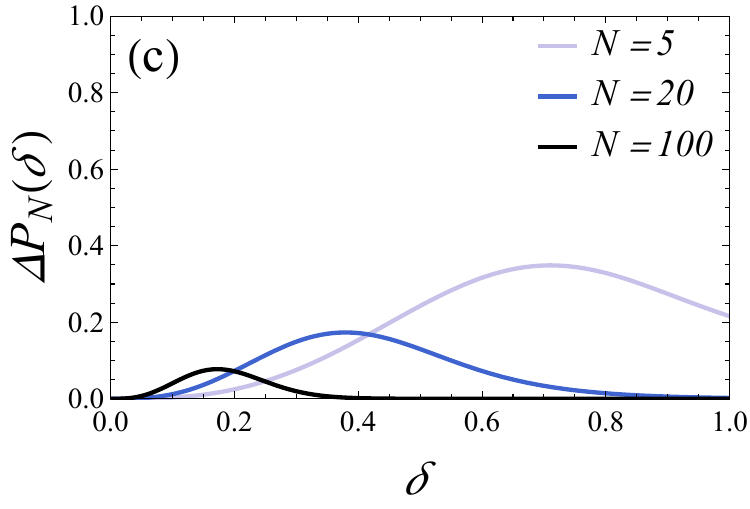}
    \end{subfigure}
    \caption{\justifying (a) The rate function of the MZM's Lyapunov exponent in the static fermionic Kitaev chain at $t=\Delta$ for different disorder strengths $W$ of the chemical potentials $\mu_n$. The disorder distribution is uniform in $[0,W]$. The topological phase transition happens at $W/4t = e$ when the minimum of $I(x)$ sits at $x=0$. (b) LDP approximation of the probability distribution of the MZM's Lyapunov exponent $L_N=x$ at $t=\Delta$ when $W/4t=1/2$ for different system sizes. (c) Asymmetric measure of the distributions in (b) as a function of distance $\delta$ from the maximum $L_\infty$ of the distributions, see \cref{eq:asym}. The dashed lines in (a)-(b) are at the thermodynamic value $x=L_\infty$.}
    \label{fig:staticFKC}
\end{figure*}

\textit{Large deviations theory}.---The theory of large deviations characterizes how the probability \mbox{$P(L_N\in[a,b])$} of a random variable, here the Lyapunov exponent $L_N$ of an edge mode, scales with sample size $N$ as $N\to\infty$. These are asymptotic scaling laws of the form $P(L_N\in[a,b]) \sim e^{-N\min_{x\in[a,b]}I(x)}$ \cite{Dembo1998}, or $P(L_N=x) \sim e^{-NI(x)}$ when $a=b=x$, known as large deviation principles (LDPs). We use the symbol $\sim$ since the LDP only captures the exponential decay rate of fluctuations away from the mean $L_\infty$ as the system size $N$ approaches infinity \footnote{We remark that it is possible to use a Bahadur-Rao approximation \cite{Bahadur1960} to obtain polynomial prefactors and sharpen the LDP estimate, but this is unnecessary for our purposes.}. The central object of this framework is the rate function $I(x)\geq 0$. By the G\"artner-Ellis theorem \cite{Dembo1998,Touchette2009}, $I(x)$ can be computed from the Fenchel-Legendre transform of the cumulant generating function (CGF) of $L_N$ given by $\Lambda(s) = \lim_{N\to\infty}\frac{1}{N}\ln \expval{e^{sNL_N}}$. That is,
\begin{align}
    \label{eq:FL_transform}
    I(x) = \sup_{s}[xs-\Lambda(s)], \quad \Lambda(s)= \sup_x[xs-I(x)],
\end{align}
where $\sup$ is the supremum. A key consequence of the Legendre duality \cref{eq:FL_transform} is that slopes $x=\Lambda'(s)$ are exactly mapped to slopes $s=I'(x)$ \cite{Touchette2009}. 
This Legendre duality between $I(x)$ and $\Lambda(s)$ is akin to the relationship between the canonical free energy and microcanonical entropy in statistical mechanics \cite{Ellis1985}. In this formal analogy, $L_N$ plays the role of energy and $\Lambda(s)$ the associated free energy at the inverse temperature $s$. Just as entropy is usually concave, the rate function is often convex, which is the case here. One of the main properties of the CGF is that $\Lambda'(0) = \expval{L_N} = L_\infty$. By Legendre duality, $I'(L_\infty)=0$. In fact, $x=L_\infty$ is the unique global minimum of $I(x)$ with $I(L_\infty) =0$. Intuitively, this is simply because $x=L_\infty$ is the typical (disorder-averaged) value of the Lyapunov exponent, recovered in the thermodynamic limit with probability $P(L_N=L_\infty)\to 1$. Below, we show that there is an asymmetry in the fluctuations of Lyapunov exponents for topological edge modes. This asymmetry is seen explicitly from the tails of the rate function: the left tail is asymptotically linear with $I'(x)\to s^*$ as $x\to -\infty$ for some finite constant $s^*$, while the right tail is infinitely steep with $I'(x)\to\infty$ as $x\to x^*$ for some constant $x^*$.

\textit{Static fermionic Kitaev chain}.---Majorana zero modes (MZMs) are topologically-protected edge modes in \textit{p}-wave superconductors \cite{Read2010,Beenakker2014} that have been suggested to exist in semiconductor heterostructures \cite{Sau2010,Alicea2010,Lutchyn2010,Oreg2010}. Their energies in finite systems are exponentially small, proportional to the left/right modes' overlap \cite{Kitaev2001}. Due to the global nature of topology, MZMs are robust against local perturbations, making them promising building blocks for performing fault-tolerant quantum computations \cite{Bravyi2002,Kitaev2003,Nayak2008,Oreg2010,Sarma2015}.
The Kitaev model \cite{Kitaev2001} simply demonstrates this on a spinless fermionic chain: 
\begin{align}
    \label{eq:Hstatic}
    \begin{split}
    H&= -\sum_{n=1}^N \mu_n c_n^\dagger c_n-t\sum_{n=1}^{N-1} \Big(c_{n+1}^\dagger c_n +c_n^\dagger c_{n+1}\Big)\\
    &\qquad +\Delta\sum_{n=1}^{N-1}\Big( c_{n+1}^\dagger c_n^\dagger+c_nc_{n+1}\Big)
    \end{split}
\end{align}
with the annihilation/creations operators $c_n,c_n^\dagger$ on site $n$. We take the chemical potentials $\mu_n$ to be independent and identically distributed (\textit{iid}) random variables, keeping uniform the hopping amplitude $t>0$ and the superconducting pairing $\Delta>0$. For a chain with uniform chemical potential $\mu_n=\mu$, the topological phase with Majorana zero-energy modes corresponds to $|\mu|<t$ and the trivial phase to $|\mu|>t$. Writing the Hamiltonian in terms of the Majorana operators $\gamma_{n}^A = c_n + c_n^\dagger$ and $\gamma_{n}^B = -i(c_n-c_n^\dagger)$,
we obtain the zero-energy dynamics of the Majorana wavefunctions $\psi^{A/B}=(\psi^{A/B}_1,\dots,\psi_N^{A/B})$ as
$[\gamma_{n}^A,H]=[\gamma_{n}^B,H]=0$, hence
\begin{align}   
    \label{eq:MZMdynamics}
    (t+\Delta)\psi_{n+1}^A + (t-\Delta)\psi_{n-1}^A&=-\mu_n\psi_{n}^A \\
    (t-\Delta)\psi_{n+1}^B + (t+\Delta)\psi_{n-1}^B&=-\mu_n\psi_{n}^B. \label{eq:MZMdynamics2}
\end{align}
The amplitudes $\psi_n^{A}$ thus satisfy 
\begin{align}
    \label{eq:transferMat}
    \begin{pmatrix}
        \psi_{n+1}^A \\
        \psi_{n}^A 
    \end{pmatrix} &= T^A_n\begin{pmatrix}
        \psi_{n}^A \\
        \psi_{n-1}^A
    \end{pmatrix}, \quad  T^A_n = \begin{pmatrix}
        \frac{-\mu_n}{t+\Delta} & -\frac{t-\Delta}{t+\Delta} \\ 1 & 0
    \end{pmatrix}
\end{align}
and the same for the transfer matrices $T_n^B$ for the other MZM edge mode $\psi_n^B$, with a change in the sign of $\Delta$. We first consider the case $t=\Delta$. The dominant eigenvalue of the product $T^A_N\cdots T^A_1$ is $e^{NL_N}$ where $L_N = \frac{1}{N}\sum_{n=1}^N X_n$ is the finite-size Lyapunov exponent and $X_n = \ln \left|\frac{\mu_n}{2t}\right|$.
\begin{figure*}[t]
    \centering
    \begin{subfigure}{0.44\textwidth}
    \centering
    \includegraphics[width=\textwidth]{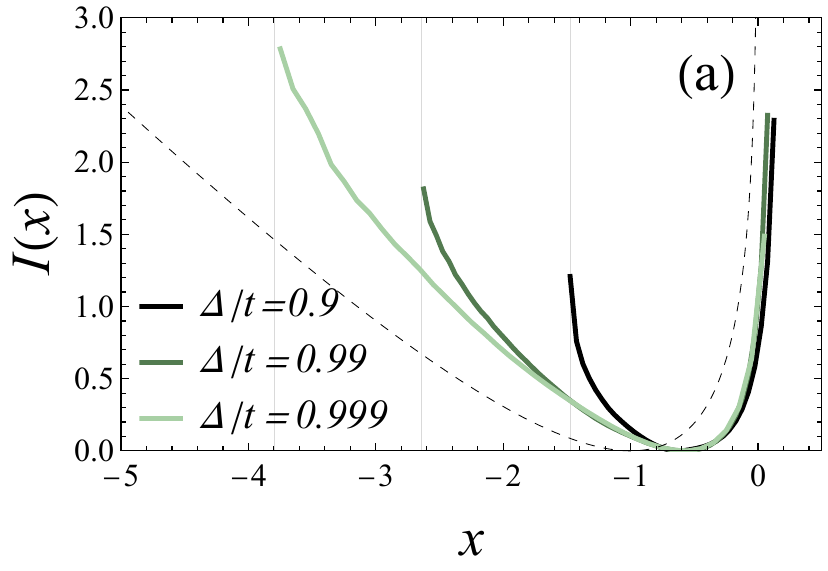}
    \end{subfigure}
    \begin{subfigure}{0.45\textwidth}
    \centering
    \includegraphics[width=\textwidth]{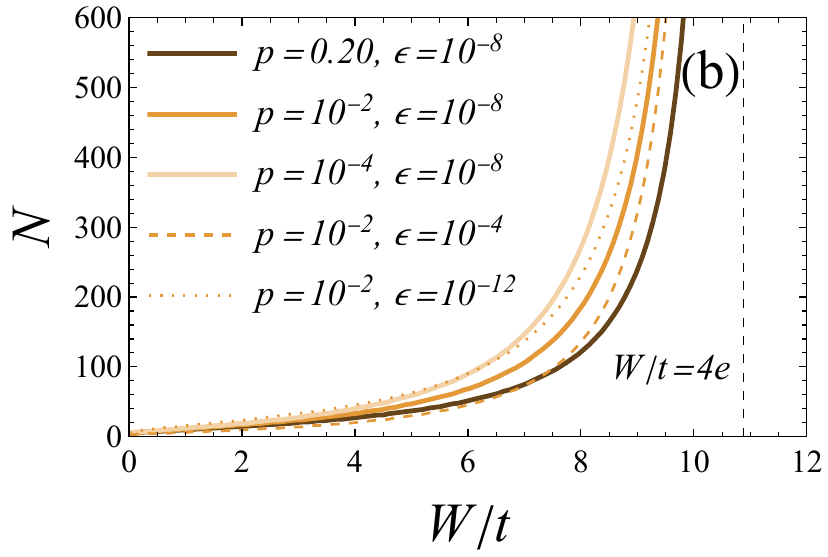}
    \end{subfigure}
    \caption{\justifying (a) Numerical estimation of the rate function governing the fluctuations of the MZM's Lyapunov exponent for different values of $\Delta/t$. To estimate $I(x)$, we generate $10^7$ disordered Kitaev chains of $N=5$ sites with uniformly sampled chemical potentials $\mu_n\in[-W/2,W/2]$ and $W/t=4$. We extract the Lyapunov exponent of each chain from the largest singular value $\sigma_1$ of the matrix product $T_5^A\cdots T_1^A$ (where $T_n^A$ is defined in \cref{eq:transferMat}) by setting $L_N = \frac{1}{N}\ln (\sigma_1)$. We bin the values of $L_N$ and normalize the resulting histogram. We estimate the value of the rate function in each of these bins $x_i$ of height $P(x_i)$ as $I(x_i) = -\frac{1}{N}\ln P(x_i)-\alpha$, where we take $\alpha \equiv \min_i I(x_i)$ to make the curve touch the $x$-axis. Note that $\alpha\to 0$ as $N\to\infty$ by the LDP. The gray vertical lines are at $x=\ln \sqrt{c}$ for the different values of $\Delta/t$. The dashed curve corresponds to the exact rate function of \cref{eq:rateFunc_static} when $\Delta/t=1$. (b) Minimum system size $N$ necessary to satisfy the condition $P(\delta \epsilon_N>\epsilon)\lesssim p$ expressed in \cref{eq:probGap_staticFKC} using the large deviation principle when $\Delta/t=1$. The vertical dashed line is the disorder value $W$ at which there is a topological phase transition in the thermodynamic limit.}
    \label{fig:extension}
\end{figure*}
Since $L_N$ is a sample mean of \textit{iid} random variables, it satisfies the LDP by Cramér's theorem \cite{Dembo1998}. Moreover, direct computation yields $\ln \expval{e^{sNL_N}}=N\ln\expval{e^{sX_n}}$ for any $N$, meaning the CGF can simply be written as $\Lambda(s) = \ln \expval{\left|\frac{\mu_n}{2t}\right|^s}$. We remark that $\Lambda(s)$ does not depend on the specific $n$ label, as the $\mu_1,\dots,\mu_N$ have the same distribution. Assuming a uniform distribution $p(\mu_n)=1/W$ for $\mu_n\in[-W/2,W/2]$,  
\begin{align}
    \expval{\left|\frac{\mu_n}{2t}\right|^s} &= \int_{-W/2}^{W/2}\frac{d\mu_n}{W}\left|\frac{\mu_n}{2t}\right|^s = \left(\frac{W}{4t}\right)^s\frac{1}{s+1}
\end{align}
for $s> -1$. Therefore, the CGF is
\begin{align}
    \Lambda(s) &= s\ln (W/4t) - \ln(s+1), \quad s>-1.
\end{align}
By definition of the CGF, the typical Lyapunov exponent is $L_\infty= \Lambda'(0)=\ln(W/4te)$. It is negative in the topological phase $W<4te$ and positive in the trivial phase. To compute the rate function from $\Lambda(s)$, we use \cref{eq:FL_transform}. Extremizing $sx-\Lambda(s)$ over $s> -1$, we find that the maximum happens at $s^* = (\ln(W/4t)-x)^{-1}-1$. Therefore, $I(x) = xs^* - \Lambda(s^*)$ gives
\begin{align}
    \label{eq:rateFunc_static}
    I(x) &= \ln(W/4te)-x-\ln(\ln(W/4t)-x).
\end{align}
\noindent The rate function has two distinct tail behaviors and shifts horizontally when changing the disorder strength $W$, see \cref{fig:staticFKC}~(a). In particular, its left tail is asymptotically linear, that is $I'(x)\to -1$ as $x\to-\infty$, while the right tail is infinitely steep with $I'(x) \to \infty$ as $x\to \ln(W/4t)$. In fact, the asymmetry about the minimum $x=L_\infty$ holds in general: fluctuations of the finite-size Lyapunov exponent $L_N$ above the thermodynamic value $L_\infty$ are exponentially less likely than those below. Specifically, for any $\delta>0$,
\begin{align}
    \label{eq:broken_symmetry}
    I(L_\infty-\delta) < I(L_\infty+\delta).
\end{align}
This asymmetry is one of our central results. By the LDP, this asymmetry carries over to probability distribution of the Lyapunov exponent, see \cref{fig:staticFKC}~(b). To quantify the distribution's asymmetry, we introduce the quantity $\Delta P_N(\delta) \equiv P(L_N=L_\infty-\delta) - P(L_N=L_\infty+\delta)$. It expresses the difference between the probability that the MZM is more strongly localized on the edge than the thermodynamic value $L_\infty$ and more weakly localized. Using the LDP,
\begin{align}
    \label{eq:asym}
    \Delta P_N(\delta) &\sim e^{-NI(L_\infty+\delta)}(e^{N[I(L_\infty+\delta)-I(L_\infty-\delta)]}-1).
\end{align}
We plot this quantity for various system sizes in \cref{fig:staticFKC}~(c). These results showcase that the MZM is exponentially more likely to be more localized on the left edge due to fluctuations than more extended, furthering its robustness to disorder. Since the transfer matrices $T_n^A$ and $T_n^B$ obtained from \cref{eq:MZMdynamics,eq:MZMdynamics2} are the same up to a sign in $\Delta$, the same is true of the MZM localized on the right edge---it is exponentially more likely to be more localized on the right edge than more extended. We observe that this asymmetry is still perceptible in relatively large chains with $N=100$ but is more prominent for small system sizes, see \cref{fig:staticFKC}~(b)-(c). This is expected from the central limit theorem, which dictates that the probability distribution $P(L_N=x)$ of the sample mean $L_N$ converges to a Gaussian distribution at a rate $1/\sqrt{N}$ \cite{Berry,Esseen}. In the general case when $t>\Delta$, the LDP still applies \cite{LePage1982,Xiao2020LDP} but the transfer matrices $T^A_n$ are no longer triangular and the Lyapunov exponent $L_N$ cannot be expressed as a sample average of $X_n=\ln|\mu_n/2t|$. What remains true here is that $L_N$ is bounded above since $\mu_n\leq W$. 
The right tail of the rate function is hence still infinitely steep. However, the left tail is now also infinitely steep since $L_N$ becomes bounded below regardless of the disorder distribution: $L_N \geq \frac{1}{2N}\ln |\det(T^A_N\cdots T^A_1)| =\ln \sqrt{c}$ with $c = \frac{t-\Delta}{t+\Delta}$. Therefore, $P(L_N<\ln\sqrt{c})=0$ hence $I(L_N=\ln \sqrt{c})=\infty$. Since the lower bound is $\ln\sqrt{c}$, the asymmetry in the fluctuations of $L_N$ is suppressed exponentially fast as $\Delta/t$ is decreased from 1, see \cref{fig:extension}~(a).

The asymmetric features of the rate function are \textit{not} a consequence of the topological phase, since they hold even when the typical Lyapunov exponent is positive, that is, when the MZM is Anderson localized by disorder, see \cref{fig:staticFKC}~(a). This can be explicitly seen from the skewness coefficient $\gamma = \Lambda'''(0)/[\Lambda''(0)]^{3/2}=-2$ of the Lyapunov exponent's distribution, which is independent of the model's parameters $t$ and $W$. In fact, the asymptotically linear left tail and the infinitely steep right tail are features of a certain class of disorder distributions containing the uniform distribution $p(\mu_n)=1/W$, which we chose for analytical convenience. The tails of the rate function found here only depend on two features of $p(\mu_n)$. First, its support needs be bounded, such that $|\mu_n|<W<\infty$. This condition imposes the infinitely steep right tail, since $P(L_N>\ln({W}/{4t})) =0$ implies $I(\ln({W}/{4t}))=\infty$ by the LDP. To have the asymptotically linear left tail, $P(L_N=x)$ needs to decay exponentially as $x\to -\infty$. Thus, the Lyapunov exponent needs to be unbounded from below. When disorder is absent with $\mu_n=0$ for all $n$, we note that $L_N=-\infty$. It is hence sufficient for the distribution $p(\mu_n)$ to be nonzero and well-behaved near $\mu_n=0$. Since the chemical potential on a given site in experiments cannot be unbounded, we anticipate that observing this tail asymmetry can be feasible as long as arbitrarily small values of $\mu_n$ are probable. See our supplemental material for details. 

\textit{Finite energy splitting.}---
Due to the finite size of the chain, $N$, the MZMs have an overlap $\sim e^{NL_N}$ \cite{Kitaev2001} which makes their degenerate eigenenergies split away from zero by an amount that scales like $\delta\epsilon_N \sim e^{NL_N}$. Strongly localized MZMs are accompanied with exponentially small energies and a large bulk gap. The quantity $\delta \epsilon_N$ thus provides a natural figure of merit to characterize the performance of a topological quantum computing device. Suppose that a computational protocol requires $\delta \epsilon_N < \epsilon$ with probability $1-p$. Our framework allows us to find the minimum system size $N$ such that $P(\delta\epsilon_N >\epsilon)\lesssim p$ at the level of large deviations, beyond the strong disorder regime and central limit approximations where the energy splitting has a log-normal distribution \cite{Brouwer2011}. Since we want $p\ll 1$, it suffices to treat the case when the disorder-averaged splitting $e^{NL_\infty}$ is below the threshold $\epsilon$. We want the probability that $\delta\epsilon_N$ exceeds that threshold to be less than $p$. The LDP directly gives $P(\delta\epsilon_N > \epsilon) \sim \exp(-N \min_{x>(1/N)\ln \epsilon}I(x))$ and using \cref{eq:rateFunc_static}, we can numerically solve
\begin{align}
    \label{eq:probGap_staticFKC}
    \left(\frac{4te}{W}\left[\ln(W/4t)-\frac{1}{N}\ln \epsilon\right]\right)^N \epsilon<p
\end{align}
for the minimum system size $N$, see \cref{fig:extension}~(b). We observe that the minimum system size grows superexponentially with the disorder strength $W$ when it approaches the topological phase transition $W=4te$. While the minimum value $N$ that satisfies the above equation may not be exact, it should accurately capture the system size dependence on $p$, $\epsilon$ and disorder strength $W$. Such probability scalings may prove useful for the implementation of small Kitaev chains ($N\sim 10-100$), and specifically for the so-called poor man's Majoranas (PMMs) hosted in quantum dot arrays operating at $\Delta=t$ \cite{Sau2012,Leijnse2012,Dvir2023,tenHaaf2024,Zatelli2024,Nitsch2025}. While these systems are too small to endow topological protection to the edge states, the rate function asymmetry provides protection from fluctuations occurring due to voltage noise. 




\begin{figure*}[t!]
    \centering
    \raisebox{0.1cm}{\begin{subfigure}{0.32\textwidth}
        \centering
        \includegraphics[width=\textwidth]{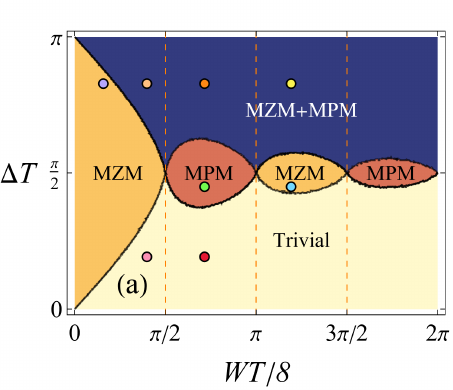}
    \end{subfigure}} 
    \raisebox{0.0cm}{
    \begin{subfigure}{0.325\textwidth}
        \centering
        \includegraphics[width=\textwidth]{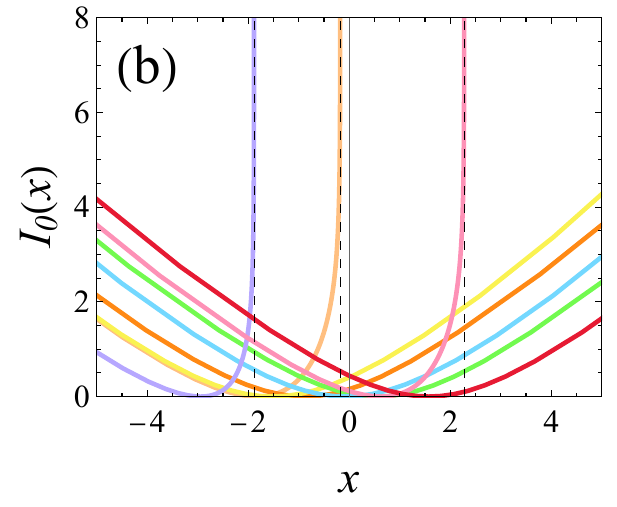}
    \end{subfigure}}
    \raisebox{0.0cm}{\begin{subfigure}{0.325\textwidth}
        \centering
        \includegraphics[width=\textwidth]{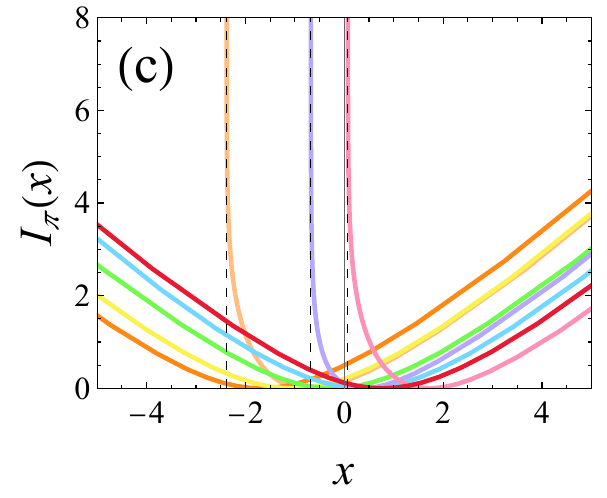}
    \end{subfigure}}
    \caption{\justifying (a) Topological phase diagram for the driven Kitaev chain at $t=\Delta$ with disordered chemical potentials uniformly sampled $\mu_n\in[-W/2,W/2]$ and \textit{infinite} system size $N\to\infty$. The chain hosts an MZM (MPM) when the Lyapunov exponent $L_\infty^{(0)}<0$ ($L_\infty^{(\pi)}<0$). The topological phase is thus determined by looking at which of $L_\infty^{(0/\pi)}$ in \cref{eq:MZM_MPM_Lyap} is negative. For numerical purposes, we took a system size of $N=10^5$ sites. The orange vertical dashed lines correspond to $WT/8=n\pi/2$. (b)--(c) The rate functions of the Lyapunov exponents computed numerically at parameter values corresponding to the colored points in (a) for the MZM (b) and the MPM (c). When $WT/8>\pi/2$, the rate functions $I_{(0/\pi)}$ lose their infinitely steep tail and become symmetric about their minimum. The vertical dashed lines are at $x=\ln[\cot(\Delta T/2)\tan(WT/8)]$ for the MZM and $x=\ln[\cot(\Delta T/2)\cot(WT/8)]$ the MPM.}
    \label{fig:drivenKC}
\end{figure*}

\textit{Driven fermionic Kitaev chain}.---By driving topological systems periodically, one may access a richer phase diagram than that obtained at equilibrium \cite{Kitagawa2010,Rudner2013,Khemani2016,Kishony2025}. Moreover, the ability to control the topological phase with an external drive parameter is a particularly desirable experimental advantage, since accessing other system parameters is generically harder. In periodically driven systems, continuous time-translation symmetry is reduced to a discrete symmetry given by the driving period $T$, forcing eigenenergies $\epsilon$ to be defined modulo $\Omega \equiv 2\pi/T$. Consequently, these quasienergies lie in the interval $\epsilon T\in(-\pi,\pi]$ and the associated eigenstates are stroboscopic. In the driven fermionic Kitaev chain, the system may host MZMs as it does in equilibrium, but it may additionally host Majorana $\pi$ modes (MPMs) \cite{Jiang2011,Lindner2011,Ling2024}. These latter states are similar to MZMs, in that they are exponentially localized on the edges of the system. However, they instead have a quasienergy of $\epsilon T=\pi$. To model the driven fermionic Kitaev chain, we consider the stroboscopic Floquet time evolution given by $U(T) = e^{-iH_2T/2}e^{-iH_1T/2}$ over one period $T$, with 
\begin{align}
    H_1 &= \sum_{n=1}^{N-1}\Big(\Delta c_{n+1}^\dagger c_n^\dagger -tc_{n+1}^\dagger c_n + \text{H.c}\Big),\\
    H_2 &= -\sum_{n=1}^N \mu_n c_{n}
    ^\dagger c_n.
\end{align}
Surprisingly, a closed form of the Lyapunov exponent of the MZM and MPM with disordered $\mu_n$'s can be computed analytically for $t=\Delta$ using the transfer matrix formalism \cite{Ling2024}. For the left-edge Majorana modes, 
\begin{align}
    \label{eq:MZM_MPM_Lyap}
    L_N^{(0/\pi)} &= \ln |\cot (\Delta T/2)| \pm \frac{1}{N}\sum_{n=1}^N \ln |\tan (\mu_n T/4)|.
\end{align}
We denote the typical (disorder-averaged, or equivalently, thermodynamic) values as $L_\infty^{(0/\pi)}\equiv \big\langle L_N^{(0/\pi)}\big\rangle$. As in the static case, we choose the chemical potentials $\mu_n\in[-W/2,W/2]$ to be \textit{iid} uniform random variables. Depending on the disorder strength $W$, the driving period $T$ and the superconducting pairing $\Delta$, the system may host either an MZM (when $L_N^{(0)}<0$), an MPM (when $L_N^{(\pi)}<0$), both or neither. Thus, one strategy to achieve large negative values of $L_N^{(0/\pi)}$ is to operate near $\Delta T/2 = \pi/2$, since $\ln\cot(\pi/2)=-\infty$. Importantly, for a large region of parameter space, disorder may never be enough to drive a topological phase transition in the \textit{infinite} system. This is a notable feature of \cref{fig:drivenKC}~(a)---which is a reproduction of the phase diagram found in Ref.~\cite{Ling2024} for our specific uniform disorder distribution---since most of the phase diagram is occupied by a phase that is nontrivial.

Since $\mu_n$'s are \textit{iid}, the Lyapunov exponents $L_N^{(0/\pi)}$ satisfy the LDP $P(L_N^{(0/\pi)}=x)\sim e^{-NI_{0/\pi}(x)}$ by Cramér's theorem \cite{Touchette2009}. Consider first $L_N^{(0)}$. For $I_0(x)$ to have an asymptotically linear left tail and an infinitely steep right tail, we need $L_N^{(0)}$ to be bounded from above and unbounded from below, that is $L_N^{(0)}\in(-\infty,x^*]$. Looking at \cref{eq:MZM_MPM_Lyap}, we want $\mu_n$ to take values that have $0\leq |\tan(\mu_nT/4)|<\infty$. One way to achieve this is again by taking a disorder distribution that is nonzero and well-behaved near $\mu_n=0$ and by taking the disorder strength $WT/8<\pi/2$. Specifically, the uniform distribution $\mu_n\in[-W/2,W/2]$ with $W/2<\Omega$ falls into this category, giving asymmetric tails for the rate function of the MZM that favor edge localization. The infinitely steep right tail of the rate function (that is, $I_0'(x)\to\infty$ as $x\to x^*$) then happens when $\mu_n=W$ for all $n=1,\dots,N$ in \cref{eq:MZM_MPM_Lyap}, with the vertical asymptote at
\begin{align}
    x^* = \ln \cot(\Delta T/2) +\ln\tan(WT/4).
\end{align}
The asymptotically linear left tail here is $I_0'(x)\to -1$ as $x\to -\infty$, see supplemental material. As we cannot obtain a closed form for the rate function $I_0(x)$, we compute it numerically and plot it in \cref{fig:drivenKC}~(b) for various parameter values. The tail asymmetry gives rise to a global asymmetry, expressed again by \cref{eq:broken_symmetry}. Regardless of the value $\Delta T/2$, we observe that the rate function is only asymmetric when $WT/8<\pi/2$, confirming the above intuition. When $WT/8>\pi/2$, it becomes symmetric since $\tan(\mu_nT/4)$ is no longer bounded above. This abrupt transitions occurs precisely when $W/2$ is equal to the driving frequency, $W_c/2=2\pi/T=\Omega$. See our supplemental material for proofs. We again emphasize that the presence of the asymmetry hinges on $W/2<\Omega$ and \textit{not} on the topological phase, see \cref{fig:drivenKC}. 

\Cref{eq:MZM_MPM_Lyap} shows that the Lyapunov exponents satisfy the reflection symmetry $L_N^{(0)} -\ln\cot(\Delta T/2) = -[L_N^{(\pi)}-\ln\cot(\Delta T/2)]$ which carries over directly to their rate functions via the LDP and gives
\begin{align}
    I_0(\ln \cot (\Delta T/2)+x) =I_{\pi}(\ln \cot (\Delta T/2)-x).
\end{align}
This relation identifies the right and left tails of $I_{0}(x)$ with the left and right tails of $I_{\pi}(x)$, respectively. Therefore, when $I_0(x)$ is asymmetric with an asymptotically linear left tail and an infinitely steep right tail, $I_\pi(x)$ is asymmetric with an infinitely steep left tail and an asymptotically linear right tail, see \cref{fig:drivenKC}~(b)--(c). We highlight that the asymmetry we can be reversed to make the MPMs more robust than the MZMs if $|\mu_n|\in[\Omega,2\Omega)$ such that $0<|\tan(\mu_nT/4)|\leq \infty$. What remains true is that the asymmetry in the driven system only enhances the robustness of \textit{one} Majorana edge mode at a time.

\textit{Discussion.}---Superconducting systems such as the Kitaev chain can encode fault-tolerant topological qubits due to the natural robustness of topological edge modes to disorder. In addition to such global robustness, we showed that the rate function of the Lyapunov exponent for the Majorana edge modes can be asymmetric---favoring stronger edge localization over more extended profiles. In the static Kitaev chain, the asymmetry is independent of the disorder strength $W$. However, this is not the case in the driven Kitaev chain, which can host MZM and MPM. Indeed, the rate functions governing the Lyapunov exponents of the MZM and MPM undergo an abrupt change from asymmetric to symmetric when $W/2>\Omega$. Furthermore, when the rate functions are asymmetric, only one of the MZM and MPM favors edge localization. We showed that this follows directly from a reflection symmetry of the Lyapunov exponents.

The rate function asymmetry we uncover provides the MZMs with an additional robustness to finite-size effects that is inherent to the transfer matrix structure of Lyapunov exponents. This fluctuation asymmetry applies to a broad class of disorder distributions and could be used to benchmark the performance of short-to-medium sized Kitaev chains in experiments, as well as their approach to the thermodynamic limit. Future research should verify whether the asymmetry we discussed here survives in the presence of particle interactions and if it also applies to Andreev bound states. It would also be interesting to study the impact of the large deviations we describe in this work on braiding and whether it extends to correlated disorder \cite{Truong2026}. 

Since the transfer matrices $T_n^A$ apply broadly to all one-dimensional topological lattice systems, the rate function asymmetry we discussed is not simply a feature of class D topological superconductors, but of edge modes in general one-dimensional topological sytems, both Hermitian and non-Hermitian. For example, it will be present in the Su-Schrieffer-Heeger model \cite{Su1979,Heeger1988} as well as the Hatano-Nelson model \cite{Hatano1996,Hatano1997,Hatano1998}. In disordered Hatano-Nelson models, the sign of the Lyapunov exponent is linked to the spectral winding number associated with non-Hermitian topology \cite{Fortin2025v2}. Therefore, the fluctuations of the spectral winding number between disorder realizations are also dictated by the rate function whose asymmetry favors nonzero integers.

\textit{Acknowledgments}.---We thank Suman Jyoti De for useful discussions. We acknowledge the support from the Qu\'ebec’s Minist\`ere de l’\'Economie, de l’Innovation et de l’\'Energie
(MEIE), Photonique Quantique Qu\'ebec (PQ2), Natural
Sciences and Engineering Research Council of Canada
(NSERC) [ALLRP 588334-23] and the Fonds de recherche
du Qu\'ebec (FRQNT). All authors benefit from their membership in the \href{ https://doi.org/10.69777/309032}{Regroupement qu\'eb\'ecois sur
les mat\'eriaux de pointe (RQMP)} .

\bibliography{references.bib}

\end{document}


\preprint{APS/123-QED}

\title{Supplemental Material for ``Finite and disordered Kitaev chains: a large deviations study"}

\author{Cl\'ement Fortin$^1$}
\author{Kai Wang$^{1}$}
\author{T. Pereg-Barnea$^{1}$}

\affiliation{$^1$Department of Physics,  Regroupement qu\'{e}b\'{e}cois sur les mat\'{e}riaux de pointe and McGill Quantum Center, McGill University, Montr\'eal, Qu\'ebec H3A 2T8, Canada }

\maketitle

\tableofcontents

\section{Introduction}
\noindent In the static and driven Kitaev chains subject to disordered chemical potentials $\mu_n$ on each site, the left edge mode's wavefunction can be written as $\psi_n\sim e^{nL_N}$, where $L_N$ is its Lyapunov exponent in a system of length $N$. The edge mode is localized on the left edge when $L_N<0$ (topological phase) and Anderson localized by disorder when $L_N>0$ (trivial phase). The precise value of $L_N$ depends on the specific values of $\mu_1,\dots,\mu_N$ and thus fluctuates depending on disorder realization. In the main text, we argue that the rate function $I(x)$ governing the large deviations (fluctuations) of the topological edge mode's Lyapunov exponent are asymmetric. In this document, we support this claim by rigorously deriving such asymmetry in the static and driven Kitaev chain at $t=\Delta$ for a broad class of disorder distribution. \\

\noindent The rate function $I(x)$ comes from the large deviation principle (LDP) $P(L_N=x) \sim e^{-NI(x)}$. We assume that the disordered chemical potentials $\mu_n$ are independent and identically distributed (\textit{iid}) random variables. In both the static and driven Kitaev chain, the Lyapunov exponent is then the sample mean of \textit{iid} random variables $X_1,\dots,X_N$ given by $L_N = \frac{1}{N}\sum_{n=1}^N X_n$. The LDP then holds by Cramér's theorem \cite{Touchette2009}, that is, there exists a rate function that governs the probabilities $P(L_N=x)$ as $N\to\infty$. To determine the tails of $I(x)$, we use the Gärtner-Ellis theorem \cite{Touchette2009}, which says that 
\begin{align}
    \label{eq:ratefuncLF}
    I(x) &=\sup_s\{xs-\Lambda(s)\}
\end{align}
where $\sup$ denotes the supremum and $\Lambda(s)$ is the cumulant generating function (CGF) of $L_N$, given by 
\begin{align}
    \Lambda(s) &\equiv \lim_{N\to\infty}\frac{1}{N}\log \expval{e^{sNL_N}} = \lim_{N\to\infty}\frac{1}{N}\log \expval{e^{s\sum_{n=1}^N X_n}} = \lim_{N\to\infty}\frac{1}{N}\log \prod_{n=1}^N \expval{e^{sX_n}} = \log \expval{e^{sX_n}}.
\end{align}
The decoupling of the random variables $X_n$ in the CGF is a direct consequence of the assumption that $\mu_n$ are \textit{iid}. \Cref{eq:ratefuncLF} says that the rate function is the Legendre-Fenchel transform of the CGF. Proceeding, we assume that the CGF is convex. This in turn implies the Legendre duality slope relation
\begin{align}
    I'(x) = s \iff \Lambda'(s)=x.
\end{align}
This is the key relation we use in this document to derive tail features of the rate function from those of $\Lambda(s)$.

\section{Static Kitaev chain: universality of the rate function's tails}
\noindent In the static Kitaev chain, the Lyapunov exponent at $t=\Delta$ is given by 
\begin{align}
    L_N &= \frac{1}{N}\sum_{n=1}^N X_n \quad \text{where}\quad X_n= \log \left|\frac{\mu_n}{2t}\right|.
\end{align}
where $\mu_n$ are \textit{iid} with probability density function $p(\mu_n)$. In this section of the supplemental material, we derive the asymmetry of the rate function governing large deviations of $L_N$ when $\mu_n$ is sampled in $[a,b]$ for $a\leq 0\leq b$ and $a<b$. Setting $W/2 \equiv \max\{|a|,|b|\}$, we will assume below (unless otherwise stated) that $\mu_n\in[-W/2,W/2]$. We note that this is without loss of generality as we can always enlarge the interval $[a,b]$ by putting $p(\mu_n)=0$ for all $\mu_n\in[-W/2,W/2]\backslash [a,b]$. We show that regardless of disorder strength, $I(x)$ is asymmetric in the following sense: it has an infinitely steep right tail $I'(x)\to \infty$ as $x\to x^*$ for some finite value of $x^*$, and it has an asymptotically linear left tail $I'(x)\to s^*$ as $x\to-\infty$. As emphasized in the main text, this asymmetry only depends on two features of the disorder distribution $p(\mu_n)$. Specifically, we need the upper bound $|\mu_n|<\infty$ for all $n=1,\dots,N$ and the fact that $p(\mu_n)$ behaves as a power law $p(\mu_n)\sim \mu_n^\beta$ near $\mu_n=0$, for some exponent $\beta>-1$. In this sense, we say that the asymmetry of the rate function is universal since it does not depend on the specific distribution $p(\mu_n)$. 

\subsection{Infinitely steep right tail}
\noindent The fact that $I(x)$ has an infinitely steep right tail means that 
\begin{align}
    I'(x)\to\infty \quad \text{as}\quad x\to x^*.
\end{align}
By Legendre duality, this is equivalent to
\begin{align}
    \Lambda'(s)\to x^* \quad\text{as}\quad s\to \infty.
\end{align}
Therefore, we want to show that $\Lambda(s)$ has an asymptotically linear right tail. We derive this directly from the fact that the magnitudes of the chemical potentials are upper bounded $|\mu_n|<W/2<\infty$, meaning $p(x)=0$ for any $|x|>W/2$. We proceed via a squeeze argument. First, since $s> 0$, we have
\begin{align}
    \Lambda(s) &= \log \left(\int_{-W/2}^{W/2} d\mu_n \,p(\mu_n) \left|\frac{\mu_n}{2t}\right|^s\right)\leq s\log(\frac{W}{4t}).
\end{align}
We can get a similar lower bound by taking any small $\delta>0$ and writing $c_\delta \equiv \int_{W/2-\delta}^{W/2} d\mu_n\, p(\mu_n)>0$. This way, 
\begin{align}
    \left(\int_{-W/2}^{W/2} d\mu_n \,p(\mu_n) \left|\frac{\mu_n}{2t}\right|^s\right) \geq \int_{W/2-\delta}^{W/2}d\mu_n \,p(\mu_n)\left|\frac{\mu_n}{2t}\right|^s\geq c_\delta \left( \frac{W/2-\delta}{2t}\right)^s,
\end{align} 
and hence $\log c_\delta + s\log(\frac{W/2-\delta}{2t})\leq \Lambda(s)$. Combining both inequalities and dividing by $s$, 
\begin{align}
    \frac{\log c_\delta}{s} + \log(\frac{W/2-\delta}{2t})\leq \frac{\Lambda(s)}{s} \leq \log(\frac{W}{4t}).
\end{align} 
Since $\delta>0$ can be taken arbitrarily small, we arrive at the asymptotically linear tail
\begin{align}
    \lim_{s\to\infty}\frac{\Lambda(s)}{s}=\log(\frac{W}{4t}).
\end{align}
By Legendre duality, $I'(x)\to \infty$ as $x\to x^*=\log(W/4t)$. 

\subsection{Asymptotically linear left tail}
\noindent The asymptotically linear left tail 
\begin{align}
    I'(x) \to s^* \quad \text{as}\quad x\to -\infty
\end{align}
follows from the upper bound $|\mu_n|\leq W/2$ and the additional assumption that the disorder distribution $p(\mu_n)$ is bounded by some power law near $\mu_n=0$. By Legendre duality, we want to show that this assumption leads to the infinitely steep left tail 
\begin{align}
    \Lambda'(s)\to-\infty \quad\text{as}\quad s\to s^*.
\end{align}
For simplicity, we will assume that $\mu_n\geq 0$ but we note that one can slightly modify the argument to show that it holds in the general case when $\mu_n$ is allowed to be negative. The crucial assumption we make is that $a_\beta \mu_n^\beta \leq p(\mu_n)\leq  b_\beta\mu_n^\beta$ in some small enough interval $[0,\delta]$, where $b_\beta>a_\beta>0$ and $\beta>-1$ is the exponent. 
Writing $\Lambda(s) = \log M(s)$ in terms of the moment generating function
\begin{align}
    M(s) =\int_{0}^{W/2} d\mu_n\, p(\mu_n) \left(\frac{\mu_n}{2t}\right)^s,
\end{align}
the derivative of the CGF is $\Lambda'(s) = M'(s)/M(s)$. By splitting the integral over the intervals $[0,\delta]\cup [\delta, W/2]$, we bound the moment generating function from above
\begin{align}
    0<M(s)&\leq \frac{1}{(2t)^s}\left(\int_{0}^\delta d\mu_n\,b_\beta \mu_n^{\beta+s}+\int_\delta^{W/2} d\mu_n\, p(\mu_n)\mu_n^s\right) \leq \frac{1}{(2t)^s}\left(\frac{b_\beta \delta^{\beta+s+1}}{\beta+s+1}+\delta ^s\right),
\end{align}
where the inequality $\int_\delta^{W/2} d\mu_n\, p(\mu_n)\mu_n^s \leq \delta^s$ follows since the map defined by $\mu_n \mapsto \mu_n^s$ for $s<0$ is a decreasing function of $\mu_n>0$. Turning to the derivative 
\begin{align}
    M'(s) &= \frac{1}{(2t)^s}\int_0^{W/2} d\mu_n\, p(\mu_n) \left(\frac{\mu_n}{2t}\right)^s\log (\frac{\mu_n}{2t}),
\end{align}
we assume without loss of generality that $2t>\delta>0$ so that $\log(\delta/2t)<0$. Therefore, for $\mu_n\in(0,\delta)$, we have $p(\mu_n)\log(\mu_n/2t)\leq a_\beta \mu_n^\beta \log(\mu_n/2t)<0$. Thus,
\begin{align}
    M'(s) &= \int_0^\delta d\mu_n\, p(\mu_n) \left(\frac{\mu_n}{2t}\right)^s\log (\frac{\mu_n}{2t})+\int_\delta^{W/2} d\mu_n\, p(\mu_n) \left(\frac{\mu_n}{2t}\right)^s\log (\frac{\mu_n}{2t}) \\
    &\leq \frac{1}{(2t)^s}\left(a_\beta\int_0^\delta d\mu_n\, \mu_n^{s+\beta}\log (\frac{\mu_n}{2t})+ C\right) \\
    &= \frac{1}{(2t)^s}\left(a_\beta\delta^{s+\beta+1}\left(\frac{\log(\delta/2t)}{s+\beta+1}-\frac{1}{(s+\beta+1)^2}\right) + C \right),
\end{align}
where $C>0$ is defined by
\begin{align}
    C \equiv \begin{cases}
        0 & \text{if  }0<W<4t \\
        (W/2)^s\log(W/4t) &\text{if }W>4t.
    \end{cases}
\end{align}
For $s<0$ close enough to $-(\beta+1)$ we have $M'(s)<0$ and the CGF satisfies
\begin{align}
    \Lambda'(s) \leq \frac{a_\beta\delta^{s+\beta+1}\left(\frac{\log(\delta/2t)}{s+\beta+1}-\frac{1}{(s+\beta+1)^2}\right) + C }{\frac{b_\beta \delta^{\beta+s+1}}{\beta+s+1}+\delta^s}.
\end{align}
Taking $s\to-(\beta+1)$, the right-hand side goes to $-\infty$ meaning $\Lambda'(s)\to -\infty$ as well. In turn, the rate function has an asymptotically linear left tail $I'(x)\to-(\beta+1)$ as $x\to-\infty$. 

\subsection{Examples of disorder distributions}
\noindent Here, we evaluate whether given disorder distributions $p(\mu_n)$ of the \textit{iid} random variables $X_n=\log|\mu_n/2t|$ satisfy the two conditions necessary to have an asymmetric rate function with an infinitely steep right tail and an asymptotically linear left tail, as outlined in the previous subsections.  
\begin{enumerate}
    \item In the main text, we used the uniform disorder distribution $p(\mu_n)=1/W$ over $\mu_n\in[-W/2,W/2]$. This trivially satisfies the first condition. The second condition follows by noticing that $1/W = \mu_n^0/W$, meaning $\beta=0$. As shown in the main text, the CGF is 
    \begin{align}
        \Lambda(s) = s\log(W/4t) -\log(s+1), \quad s> -1
    \end{align}
    and the rate function is 
    \begin{align}
        I(x) = \log(W/4te) -x-\log(\log(W/4t)-x).
    \end{align}
    \item Consider the truncated Gaussian distribution $p(\mu_n) = e^{-(\mu_n-\mu_c)^2/2\sigma^2}/Z$ which integrates to unity over $[a,b]$ for $Z = \int_a^b d\mu_n \,p(\mu_n)$, with mean $\mu_c$ and variance $\sigma^2<\infty$. The first condition is again trivial, and we again find $\beta=0$ near $\mu_n=0$. Thus, the rate function is asymmetric. 
    \item Consider the full Gaussian distribution $p(\mu_n)= e^{-(\mu_n-\mu_c)/2\sigma^2}$ over $\mathds{R}$ with mean $\mu_c$ and variance $\sigma^2<\infty$. Here, it is the first condition that does not hold since $\mu_n$ can be arbitrarily large. Therefore, the rate function does not have an infinitely steep right tail $I'(x)\to \infty$ at some $x\to x^*$. However, the rate function is still very asymmetric. For this special case, it is possible to find a closed form for the rate function by making use of the contraction principle \cite{Touchette2009}. To find the rate function explicitly, we first note that the probability density function associated with the sample mean $S_N= \frac{1}{N} \sum_{n=1}^N \mu_n$ of Gaussian distributions is \cite{Touchette2009} 
    \begin{align}
        I_\mu (\mu) &= \frac{(\mu-\mu_c)^2}{2\sigma^2}. 
    \end{align}
    However, we are interested in the rate function of the sample mean $L_n = \frac{1}{N}\sum_{n=1}^N X_n$ for $X_n = \log|\mu_n/2t|$. Writing $x = \log |\mu/2t|$, we have $\mu = \pm2te^x$ and the contraction principle gives 
    \begin{align}
        \label{eq:ratefuncGaussian}
        I(x) = I_\mu(2te^x) = \frac{(2te^x - |\mu_c|)^2}{2\sigma^2}.
    \end{align}
    Thus, the rate function has an asymptotically linear left tail $I'(x) \to 0$ as $x\to-\infty$, but does not have an infinitely steep right tail at some finite value $x=x^*$. Nonetheless, it does have a rapidly increasing right tail $I'(x) \approx  (te^x/\sigma)^2$ when $x\gg1$, see \cref{fig:rateFuncGaussian}.
    \begin{figure}
        \centering
        \includegraphics[width=0.7\textwidth]{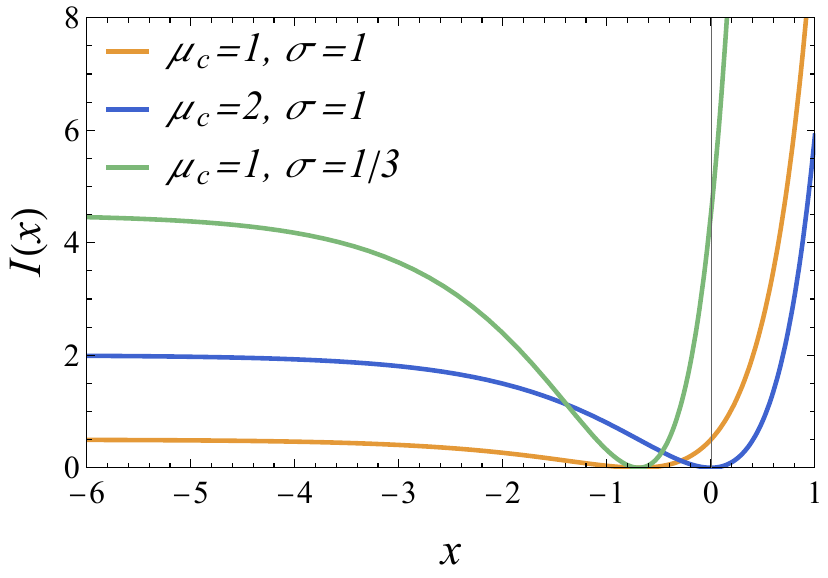}
        \caption{Rate function given in \cref{eq:ratefuncGaussian} of the Lyapunov exponent $L_N$ when the chemical potentials are distributed according to the Gaussian $\mathcal{N}(\mu_c,\sigma^2)$ for different values of mean $\mu_c$ and standard deviation $\sigma$.}
        \label{fig:rateFuncGaussian}
    \end{figure}
\end{enumerate}

\section{Driven Kitaev chain: universality of the rate function's tails}
\noindent In the Kitaev chain driven with period $T>0$ and subject to disordered chemical potentials $\mu_1,\dots,\mu_N$, the Lyapunov exponents associated with the Majorana zero mode (MZM) and the Majorana pi mode (MPM) localized on the left edge for $t=\Delta>0$ have the closed forms \cite{Ling2024}
\begin{align}
    L_N^{(0)} &= \log\cot(\Delta T/2) + \frac{1}{N}\sum_{n=1}^N\log|\tan(\mu_nT/4)|, \\
    L_N^{(\pi)} &= \log\cot(\Delta T/2) - \frac{1}{N}\sum_{n=1}^N\log |\tan(\mu_nT/4)|.
\end{align}
Since the Majorana wavefunctions are described by $\psi_n^{(0/\pi)} \sim e^{nL_N^{(0/\pi)}}$, they will be localized on the left edge when $L_N^{(0/\pi)} <0$, and Anderson localized by disorder when $L_N^{(0/\pi)}>0$. The reflection symmetry
\begin{align}
    L_N^{(0)}-\log\cot(\Delta T/2) = -[L_N^{(\pi)} - \log\cot(\Delta T/2)]
\end{align}
readily carries over to the rate functions $I_0(x)$ and $I_\pi(x)$ via the LDPs
\begin{align}
    P(L_N^{(0)}-\log\cot(\Delta T/2)=x) &= P(L_N^{(0)}=\log\cot(\Delta T/2)+x) \sim e^{-NI_0(\log\cot(\Delta T/2)+x)} \\
    P(-[L_N^{(\pi)}-\log\cot(\Delta T/2)]=x) &= P(L_N^{(\pi)}=\log\cot(\Delta T/2)-x) \sim e^{-NI_\pi(\log\cot(\Delta T/2)-x)}.
\end{align}
Specifically, $I_0(\log\cot(\Delta T/2)+x)=I_\pi(\log\cot(\Delta T/2)-x)$. The left (right) tail of the rate function $I_{0}(x)$ is then identified with the right (left) tail of the rate function $I_\pi(x)$. The fact that $I_0(x)$ and $I_\pi(x)$ have opposite asymmetry hence readily follows from their reflection symmetry about $\log\cot(\Delta T/2)$. Henceforth, we focus on determining the tail behavior of $I_0(x)$ to know that of $I_\pi(x)$. We determine the tails of $I_0(x)$ from the Legendre duality with the cumulant generating function (CGF) $\Lambda_0(s)$ of $L_N^{(0)}$. To proceed, we assume that the chemical potentials $\mu_n$ are independent and identically distributed (\textit{iid}) random variables with the probability distribution $p(\mu_n)$. In this case,
\begin{align}
    \Lambda_0(s) &= \lim_{N\to\infty} \frac{1}{N}\log \expval{e^{sNL_N^{(0)}}} = s\log \cot(\Delta T/2) + \log M_0(s), \qquad M_0(s) = \int_{-\infty}^\infty d\mu_n\, p(\mu_n)|\tan(\mu_n T/4)|^s,
\end{align}
where $M_0(s)$ is the moment generating function of the \textit{iid} random variable $X_n = \log\tan|\mu_n T/4|$ for any $n=1,\dots,N$.

\subsection{Asymptotically linear left tail}
\noindent We here show that the rate function $I_0(x)$ has the asymptotically linear left tail
\begin{align}
    I_0'(x) \to -1 \quad\text{as}\quad x\to-\infty.
\end{align}
Legendre duality dictates that this is equivalent to 
\begin{align}
    \label{eq:driven_left_tail}
    \Lambda_0'(s) \to -\infty \quad\text{as}\quad s\to -1.
\end{align}
This will follow from the fact $|\tan(\mu_n T/4)|^s \approx |\mu_nT/4|^s$ near $\mu_n=0$, which is not integrable when $s\geq 1$. In contrast to the static Kitaev chain, there is nothing special about the value $\mu_n=0$ since the tangent function has many zeros on the real line. The only thing necessary is that $p(\mu_n)$ behaves like a power law near a zero of $\tan(\mu_n T/4)$, that is, near $\mu_n=4k\pi/T$ for any integer $k$. We thus limit ourselves to the case $W<8\pi/T$. Additionally, we will assume for simplicity that $\mu_n\in[0,W/2]$ (nonnegative), but we note that the argument can be generalized to the case where $\mu_n$ takes negative values as well because we only care about the integral of the absolute value $|\tan(\mu_nT/4)|$.\\

\noindent We now show that \cref{eq:driven_left_tail} follows from the assumptions that (1) $|\mu_n|\leq W/2$ and that (2) the probability density function is bounded above and below by some power law $a_\beta \mu_n^\beta \leq p(\mu_n)\leq  b_\beta\mu_n^\beta$ in some small enough interval $\mu_n\in[0,\delta]$, where $b_\beta>a_\beta>0$ and $\beta>-1$ is the exponent. The derivative of the CGF is 
\begin{align}
    \Lambda'_0(s) = \log\cot(\Delta T/2) + M_0'(s)/M_0(s), \qquad M'_0(s) &= \int_{0}^{W/2} d\mu_n \,p(\mu_n)|\tan(\mu_nT/4)|^s\log|\tan( \mu_nT/4)|.
\end{align}
Splitting the integral over the intervals $[0,\delta]\cup[\delta,W/2]$,
\begin{align}
    0<M_0(s) \leq \left(\frac{T}{4}\right)^s\int_0^{\delta} d\mu_n\, b_\beta \mu_n^{\beta+s}+\int_\delta^{W/2} d\mu_n\, p(\mu_n) |\tan(\mu_n T/4)|^s \leq \left(\frac{T}{4}\right)^s\frac{b_\beta \delta ^{\beta+s+1}}{\beta+s+1}+ \tan^s(\delta T/4)
\end{align}
where the upper bound of the integral over $(\delta,W/2)$ follows from the fact that we are looking at the left tail $s<0$ and that $\delta>0$ is taken arbitrarily small. We now turn to the derivative $M_0'(s)$. Without loss of generality, take $\delta>0$ such that $\delta<\pi/T$. We split the derivative
\begin{align}
    M_0'(s) &= \int_0^\delta d\mu_n \, p(\mu_n) |\tan(\mu_n T/4)|^s \log|\tan(\mu_n T/4)| + \int_\delta^{W/2} d\mu_n \, p(\mu_n) |\tan(\mu_n T/4)|^s \log|\tan(\mu_n T/4)|.
\end{align}
We want to find an upper bound for $M'_0(s)$ when $s<0$. We start with the first integral over $(0,\delta)$. Let $g_s(t) \equiv t^s\log t$, which has $g_s'(t)>0$ on $t\in(0,1)$ when $s<0$. The function $g_s(t)$ is thus strictly increasing when $t\in(0,1)$. Since $\log \tan(\delta T/4)<0$, then $g_s(\tan(\mu_nT/4)) <0$ for all $\mu_n\in(0,\delta)$. Additionally, since the function $\tan(x)/x$ is strictly increasing on $(0,\delta)$, we have 
\begin{align}
    \frac{\tan(\mu_n T/4)}{\mu_n T/4}\leq \frac{\tan(\delta T/4)}{\delta T/4}\equiv \kappa\implies \tan(\mu_n T/4) \leq \kappa \mu_n T/4.
\end{align}
Therefore, the first integral in the derivative $M'_0(s)$ has
\begin{align}
    \int_0^\delta d\mu_n \, p(\mu_n) \tan^s(\mu_n T/4) \log\tan(\mu_n T/4) &\leq a_\beta \int_0^\delta d\mu_n \,\mu_n^\beta g(\tan(\mu_nT/4)) \\
    &\leq a_\beta \int_0^\delta d\mu_n \, \mu_n^\beta g(\kappa\mu_n T/4) \\
    &= a_\beta(\kappa T/4)^s \int_0^\delta d\mu_n \,\mu_n^\beta\log (\kappa \mu_n T/4) \\
    &= a_\beta (\kappa T/4) ^s \delta^{\beta+s+1}\left(\frac{\log(\kappa \delta T/4)}{\beta+s+1}-\frac{1}{(\beta+s+1)^2}\right).
\end{align}
For the second integral over $(\delta,W/2)$, note that $\tan(\mu_n T/4)>0$ over $(\delta,W/2)$ (recall that we chose $W<8\pi/T$). Thus, the function $g_s(\tan(\mu_nT/4))$ is finite over $\mu_n\in[\delta,W/2]$ and $s\in[-\beta-1,0]$. We thus bound
\begin{align}
    C \equiv \max_{s\in[-\beta-1,0],\hspace{0.1cm}t\in[\delta,W/2]} g_s(t)<\infty
\end{align}
and since $\int_\delta^{W/2} d\mu_n \, p(\mu_n)\leq 1$ by normalization of the probability density function, then
\begin{align}
    M_0'(s) &= \int_0^\delta d\mu_n \, p(\mu_n) \tan^s(\mu_n T/4) \log\tan(\mu_n T/4) + \int_\delta^{W/2} d\mu_n \, p(\mu_n) \tan^s(\mu_n T/4) \log\tan(\mu_n T/4) \\
    &\leq a_\beta (\kappa T/4) ^s \delta^{\beta+s+1}\left(\frac{\log(\kappa \delta T/4)}{\beta+s+1}-\frac{1}{(\beta+s+1)^2}\right) + C.
\end{align}
Since $M_0'(s)\to -\infty$ as $s\to -(\beta+1)$, taking $s$ close enough to $-(\beta+1)$ yields $M_0'(s)<0$. The CGF then has  
\begin{align}
    \Lambda_0'(s) &\leq \frac{a_\beta (\kappa T/4) ^s \delta^{\beta+s+1}\left(\frac{\log(\kappa \delta T/4)}{\beta+s+1}-\frac{1}{(\beta+s+1)^2}\right) + C}{\left(\frac{T}{4}\right)^s\frac{b_\beta \delta ^{\beta+s+1}}{\beta+s+1}+ \tan^s(\delta T/4)}
\end{align}
with $\Lambda'_0(s)\to -\infty$ as $s\to-(\beta+1)$. By Legendre duality, the rate function $I_0(x)$ has an asymptotically linear left tail with slope $I'_0(x)\to -1$ as $x\to-\infty$. 

\subsection{The right tail}
\noindent We here show that when $|\mu_n|\leq W/2<\infty$ (that is, $p(x)>0$ if and only if $|x|<W/2$), there is an abrupt transition in the rate function's right tail when disorder strength is taken from $W<4\pi/T$ to $W>4\pi/T$. Specifically, the rate function has the infinitely steep right tail
\begin{align}
    I'_0(x) \to \infty\quad\text{as}\quad x\to \log[\cot(\Delta T/2) \tan(WT/8)]
\end{align}
when $W<4\pi/T$, and has an asymptotically linear right tail 
\begin{align}
    I'_0(x) \to 1 \quad\text{as}\quad x\to \infty
\end{align}
when $W>4\pi/T$. Similarly to the previous subsection, there is nothing special about the specific value $4\pi/T$. The only thing that matters is that the chemical potentials can take values arbitrarily close to a pole of $\tan(\mu_n T/4)$, that is, near $\mu_n=2(2k+1)\pi/T$ for any integer $k$. We thus assume for simplicity that $W<8\pi/T$. We also make the additional assumption that $\mu_n\geq 0$, but we note that the argument can be generalized to cases where $\mu_n$ takes negative values as well. \\

\noindent We first show the tail behavior when $W<4\pi /T$, meaning $WT/8< \pi/2$. By Legendre duality, it is equivalent to 
\begin{align}
    \Lambda_0'(s) \to \log[\cot(\Delta T/2) \tan(WT/8)] \quad\text{as}\quad s\to\infty.
\end{align}
Since we are looking at regimes where $s>0$, we can bound above
\begin{align}
    \log M_0(s) &= \log\int_{0}^{W/2}d\mu_n\, p(\mu_n)\tan^s\left|\frac{\mu_n T}{4}\right|  \leq s\log\tan(\frac{WT}{8}).
\end{align}
For any small $\delta\in(0,W/2)$, let $c_\delta \equiv  \int_{W/2-\delta}^{W/2} d\mu_n\, p(\mu_n)$ so that
\begin{align}
    \log M_0(s) &\geq \log\int _{W/2-\delta}^{W/2} d\mu_n \, p(\mu_n)\tan^s\left|\frac{\mu_n T}{4}\right| \geq s\log\tan(\frac{(W/2-\delta)T}{4})+\log c_\delta.
\end{align}
Combining the two inequalities and dividing by $s$, we find 
\begin{align}
    \log \cot (\frac{\Delta T}{2}) +\frac{\log c_\delta}{s} + \log\tan(\frac{(W/2-\delta)T}{4}) \leq \frac{\Lambda_0(s)}{s} \leq \log \cot (\frac{\Delta T}{2}) +\log\tan(\frac{WT}{8}).
\end{align}
Since $\delta>0$ can be taken arbitrarily small and $c_\delta\to0$ as $\delta\to0$, we find the asymptotically linear right tail 
\begin{align}
    \lim_{s\to\infty}\frac{\Lambda_0(s)}{s} &= \log \left[\cot (\frac{\Delta T}{2})\tan(\frac{WT}{8})\right]
\end{align}
or equivalently, $\Lambda_0'(s)\to \log[ \cot(\Delta T/2)\tan(WT/8)]$ as $s\to\infty$. By Legendre duality, the rate function $I_0(s)$ does indeed have an infinitely steep right tail at $x=\log[ \cot(\Delta T/2)\tan(WT/8)]$ when $W<4\pi/T$. \\

\noindent We now turn to the case $W>4\pi /T$, or equivalently $WT/8\geq \pi/2$ (and recall our assumption that $W<8\pi/T$ meaning $WT/8<\pi$). We want to show that $\Lambda'_0(s)=\log\cot(\Delta T/2) +M_0'(s)/M_0(s)\to \infty$ as $s\to 1$. The tangent in the moment generating function  
\begin{align}
    M_0(s)=\int_{0}^{W/2}d\mu_n \, p(\mu_n)|\tan(\mu_nT/4)|^s
\end{align} 
develops a singularity at $s=1$ since $|\tan(\mu_nT/4)|^s\approx |\mu_nT/4-\pi/2|^{-s}$ near $\mu_n=2\pi/T$. We want to find a positive lower bound for $M_0'(s)$ and a positive upper bound for $M_0(s)$ whose ratio goes to $\infty$ as $s$ approaches 1. We start with $M_0(s)$. Let $\delta>0$ be small and write
\begin{align}
    \label{eq:MGF_driven}
    M_0(s) &= 
 \int_{0}^{2\pi /T-\delta} d\mu_n \, p(\mu_n) |\tan(\mu_nT/4)|^s+\int_{2\pi/T-\delta}^{2\pi /T+\delta} d\mu_n \, p(\mu_n) |\tan(\mu_nT/4)|^s+\int_{2\pi/T+\delta}^{W/2} d\mu_n \, p(\mu_n) |\tan(\mu_nT/4)|^s.
\end{align}
The first and third integrals can be bounded from above 
\begin{align}
    \int_{0}^{2\pi /T-\delta} d\mu_n \, p(\mu_n) |\tan(\mu_nT/4)|^s &\leq |\tan(\pi/2-\delta T/4)|^s\\
    \int_{2\pi/T+\delta}^{W/2} d\mu_n \, p(\mu_n) |\tan(\mu_nT/4)|^s&\leq |\tan(WT/8)|^s.
\end{align}
Set $K_{\delta,s} \equiv  |\tan(\pi/2-\delta T/4)|^s+|\tan(WT/8)|^s$. The singularity at $s=1$ comes from the second integral. Note that the Laurent series of $\tan(x)$ near $x=\pi/2$ is
\begin{align}
    \tan(x) &= \frac{1}{\pi/2-x}+\frac{1}{3}(x-\pi/2) + \frac{1}{45}(x-\pi/2)^3 + \dots
\end{align}
Near $x=\pi/2$, only the first term contributes. Thus, taking $C_\delta>0$ large enough such that $|\tan(\mu_nT/4)|^{s} \leq C|\pi/2-\mu_nT/4|^{-s}$ on $[2\pi /T-\delta,2\pi /T+\delta]$,
\begin{align}
    \int_{2\pi/T-\delta}^{2\pi /T+\delta} d\mu_n \, p(\mu_n) |\tan(\mu_nT/4)|^s &\leq C_\delta\int_{2\pi /T-\delta} ^{2\pi/T+\delta} d\mu_n\,|\pi/2-\mu_n T/4|^{-s} \leq \frac{C_\delta(\delta T/4)^{1-s}}{1-s}
\end{align}
and hence $M_0(s) \leq K_{\delta,s}+\frac{C_\delta(\delta T/4)^{1-s}}{1-s}$. Turning to the derivative $M_0'(s)$, we proceed in a similar fashion by bounding it from below. We make the substitution $x=\mu_nT/4$ and $dx = Td\mu_n /4$, and we split the integral in three:
\begin{align}
    M_0'(s) &= \frac{4}{T}\int_0^{WT/8} dx\, p(4x/T)|\tan(x)|^s \log|\tan(x)| \\
    &=\frac{4}{T}\int_0^{\pi/2-\delta T/4} dx\, p(4x/T)|\tan(x)|^s \log|\tan(x)|+\frac{4}{T}\int_{\pi/2-\delta T/4}^{\pi/2+\delta T/4} dx\, p(4x/T)|\tan(x)|^s \log|\tan(x)| \\
    &\quad +\frac{4}{T}\int_{\pi/2+\delta T/4}^{WT/8} dx\, p(4x/T)|\tan(x)|^s \log|\tan(x)|.
\end{align}
The first and third integrals can be bounded by a constant independent of $s$. Again, consider the function $g_s(t) \equiv t^s\log t$ for $s\in[0,1]$ which is finite and continuous for all $t\in[m_\delta,M_\delta]$ where 
\begin{align}
    m_\delta \equiv \min_{x\in[0,\pi/2-\delta T/4]\cup [\pi/2+\delta T/4,WT/8]} |\tan x| \quad\text{and}\quad  M_\delta \equiv \max_{x\in[0,\pi/2-\delta T/4]\cup [\pi/2+\delta T/4,WT/8]} |\tan x|.
\end{align}
Setting
\begin{align}
    D_\delta \equiv \left( \min_{s\in[0,1],t\in[m_\delta,M_\delta]}g_s(t) \right)\left(\int_{[0,\pi/2-\delta T/4]\cup [\pi/2+\delta T/4,WT/8]}\frac{4dx}{T}\, p(4x/T)\right),
\end{align}
we find 
\begin{align}
    \frac{4}{T}\int_{[0,\pi/2-\delta T/4]\cup [\pi/2+\delta T/4,WT/8]} dx\, p(4x/T)|\tan(x)|^s \log|\tan(x)| \geq D_\delta.
\end{align}
For the second integral in $M_0'(s)$, take a constant $F_\delta>0$ such that $|\tan(x)|^s \log|\tan(x)| \geq -F_\delta|\pi/2-x|^{-s}\log|\pi/2-x|>0$ for all $x\in[\pi/2-\delta T/4,\pi/2+\delta T/4]$. Then
\begin{align}
    \frac{4}{T}\int_{\pi/2-\delta T/4}^{\pi/2+\delta T/4} dx\, p(4x/T)|\tan(x)|^s \log|\tan(x)| &\geq -\frac{4F_\delta }{T} \int_{\pi/2-\delta T/4}^{\pi/2 + \delta T/4}dx\, p(4x/T)\frac{\log|\pi/2-x|}{|\pi/2-x|^s}\\
    &= -\frac{4F_\delta }{T}\int_{-\delta T/4}^{\delta T/4}dx\, p(4(\pi/2-x)/T)\frac{\log |x|}{|x|^s} \\
    &= -\frac{4F_\delta}{T}\int_0^{\delta T/4}dx\, \Big[p(4(\pi/2-x)/T)+p(4(\pi/2+x)/T)\Big] x^{-s}\log x.
\end{align}
Setting 
\begin{align}
    p_{\delta,\text{min}} \equiv \min_{x\in[0,\delta T/4]}\frac{p(4(\pi/2-x)/T)+p(4(\pi/2+x)/T)}{2}>0
\end{align}
we integrate by parts to find 
\begin{align}
    -\frac{4F_\delta}{T}\int_0^{\delta T/4}dx\, \Big[p(4(\pi/2-x)/T)+p(4(\pi/2+x)/T)\Big] x^{-s}\log x &\geq -\frac{8F_\delta p_{\delta,\text{min}}}{T}\int_0^{\delta T/4} dx\, x^{-s}\log x \\
    &\geq \frac{8F_\delta p_{\delta,\text{min}}(\delta T/4)^{1-s}}{T}\left(\frac{1}{(1-s)^2}-\frac{\log(\delta T/4)}{1-s}\right).
\end{align}
Putting everything together, 
\begin{align}
    \Lambda_0'(s) &\geq \log\cot(\Delta T/2) + \frac{\frac{8F_\delta p_{\delta,\text{min}}(\delta T/4)^{1-s}}{T}\left(\frac{1}{(1-s)^2}-\frac{\log(\delta T/4)}{1-s}\right)}{K_{\delta,s}+\frac{C_\delta (\delta T/4)^{1-s}}{1-s}}\to \infty \quad \text{ as }\quad s\to 1.
\end{align}
The domain of the CGF in this case is thus restricted to $s<1$.
By Legendre duality, the rate function $I_0(x)$ now has the asymptotically linear right tail $I'_0(x)\to 1$ as $x\to\infty$. Thus, increasing disorder strength above twice the Floquet quasienergy bandwith $W_c = 4\pi/T = 2\Omega$ restores the (asymptotic) fluctuation symmetry.

\bibliography{references.bib}